\documentclass{article}
\usepackage{graphicx}  
\usepackage{amsmath}   
\usepackage[compress]{cite}
\usepackage{amssymb}   
\usepackage{bm} 
\usepackage{dcolumn}
\usepackage{color}
\usepackage{mathrsfs}
\usepackage{amsfonts}
\usepackage{varioref}
\RequirePackage[colorlinks,citecolor=blue,urlcolor=magenta,linkcolor=blue]{hyperref}
\labelformat{section}{Section #1} 
\labelformat{subsection}{Section #1} 
\labelformat{subsubsection}{Section #1}
\labelformat{subsubsubsection}{Section #1}
\labelformat{equation}{Eq.~(#1)} 
\labelformat{figure}{Fig.~#1} 
\labelformat{subfigure}{Fig.~\thefigure#1} 
\labelformat{table}{Tab.~#1} 
\labelformat{appendix}{Appendix #1}
\title{Quantum leaps of black holes: Magnifying glasses of quantum gravity
\footnote{This essay has received honorable mention in the Gravity Research Foundation 2016 Awards for Essays on Gravitation.}}
\author{Sumanta Chakraborty $^{\dagger,}$\footnote{Email: sumanta@iucaa.in;~~sumantac.physics@gmail.com}\footnote{Corresponding Author} 
and Kinjalk Lochan
\footnote{Email: kinjalk@iucaa.in;~~kinjalk.lochan@gmail.co.in}
\\
{\small{IUCAA, Post Bag 4, Ganeshkhind, Pune University Campus, Pune 411 007, India}}}
\begin{document}
  
\maketitle
\begin{abstract}
We show using simple arguments, that the conceptual triad of a {\it classical} black hole, semi-classical Hawking emission and geometry quantization is inherently, mutually incompatible. Presence of any two explicitly violates the third. We argue that geometry quantization, if realized in nature, magnifies the quantum gravity features hugely to catapult them into the realm of observational possibilities. We also explore a quantum route towards extremality of the black holes.
\end{abstract}
\newpage
Black holes, by now, are famously known to behave very much like thermal objects, possessing entropy and radiating thermally \cite{Hawking:1974sw,Bekenstein:1973ur}. Microscopically, the entropy of any physical system originates from a statistical description of the underlying micro-states. Therefore, one expects the black hole entropy as well, to originate from a micro-state counting, as and when a consistent quantum theory of gravity is known. Although we may like to believe that we are safely removed from the regime of nitty gritties of quantum gravity, there are at least two fundamental set of issues,  as argued below, which may well have been harboring  direct signatures of a quantized spacetime geometry, within the realm of observational possibilities.
\begin{itemize}

\item The thermal radiation of black holes pose challenging issues for quantum information theory. The thermal profile of Hawking radiation makes it information less. Whatever ends in black hole gets mapped into the temperature of the hole, which only cares about its mass (and angular momentum of astrophysical black holes). Once the black hole evaporates away all other information is seemingly lost forever. Quantum gravity cannot save the day, since it gets active only near the Planck Scale. By that time, Hawking radiation would destroy enough of information in initial data.

\item Is cosmic censorship a law of nature? Even if the issue of singularities in GR is resolved in quantum gravity, one can legitimately ask if it is possible to over-spin a black hole and and destroy its event horizon? Reaching extremal limit itself is thermodynamically unachievable. So, is there any \textit{supra-thermodynamic} physical process which could make the black hole extremal and beyond ? That is to say is it possible to destroy the black holes in a physical process?

\end{itemize}
We will show using a rather general set of arguments, that a theory of quantized geometry will have strong implications viz-\'{a}-viz these issues.
\paragraph{\it \textbf{``That which does not kill you, makes you stronger :"}}

Bekenstein and Mukhanov \cite{Bekenstein:1995ju} argued, taking a simple case of {\it microscopic} area quantization that if a black hole is treated as a highly excited state of an underlying quantum description, the emission spectra becomes very sparse \footnote{Recently, similar results have been obtained in time domain \cite{Gray:2015pma,Hod:2015wva} as well.}. To be precise, the allowed minimum frequency gets close to $1/M$, where semi-classically, the thermal emission spectra should have been peaking. This provided a great opportunity of observational ramifications of such quantization models, but also ran into contradiction with the semi-classical expectation and brought the scheme of such quantization in a quagmire. However, it was suggested that the demonstration was heuristic and simplistic, at best, as the black hole was taken to be a highly excited state, rather than a collection of many microscopic states and the area was taken to be integerly quantized, rather too simplistically. The general hope was that more complex classes of geometry quantization and spectrum profiles will eliminate this apparent 
discrepancy and ``save the day for quantum geometry'' viz-\'{a}-viz semi-classical physics. We argue against the viability of such hopes and show that more complex implementation of geometry quantization does not really destroy the arguments of Bekenstein and Mukhanov, but rather paves way for their return in a much more formidable form. 

\paragraph{\it \textbf{``A tussle of titans - classical, semi-classical and quantum gravity :"}} We argue, motivating and extending the ideas of Bekenstein and Mukhanov that the components of conceptual triad of
\begin{itemize}
\item (A) macroscopic black holes with {\it classical} properties, 
\item (B) the thermal character of the {\it semi-classical} Hawking radiation, and
\item (C) idea of {\it geometry quantization}
\end{itemize} 
are mutually incompatible. Realization of any two of these three ideas will rule the third one out.  As the thermal nature of Hawking radiation results in an information loss paradox, this triad can be thought of as a possible antidote against the famous AMPS triad, potentially having significant implication for the paradox \cite{Almheiri:2012rt}. We show below, how {\bf any} quantized geometry prevents a {\it Infrared Catastrophe}, masking out low frequency emission and strengthening high frequency emission of black holes, thereby making it essentially non-thermal. We show that combination of (A) and (C), not only whisks the information loss seemingly away but also significantly enhances the observational prospects of first quantum gravity signatures.
\paragraph{\it \textbf{The unbecoming of the Hawking spectrum :}}

\begin{figure}[t!]
\centering
\includegraphics[scale=0.5]{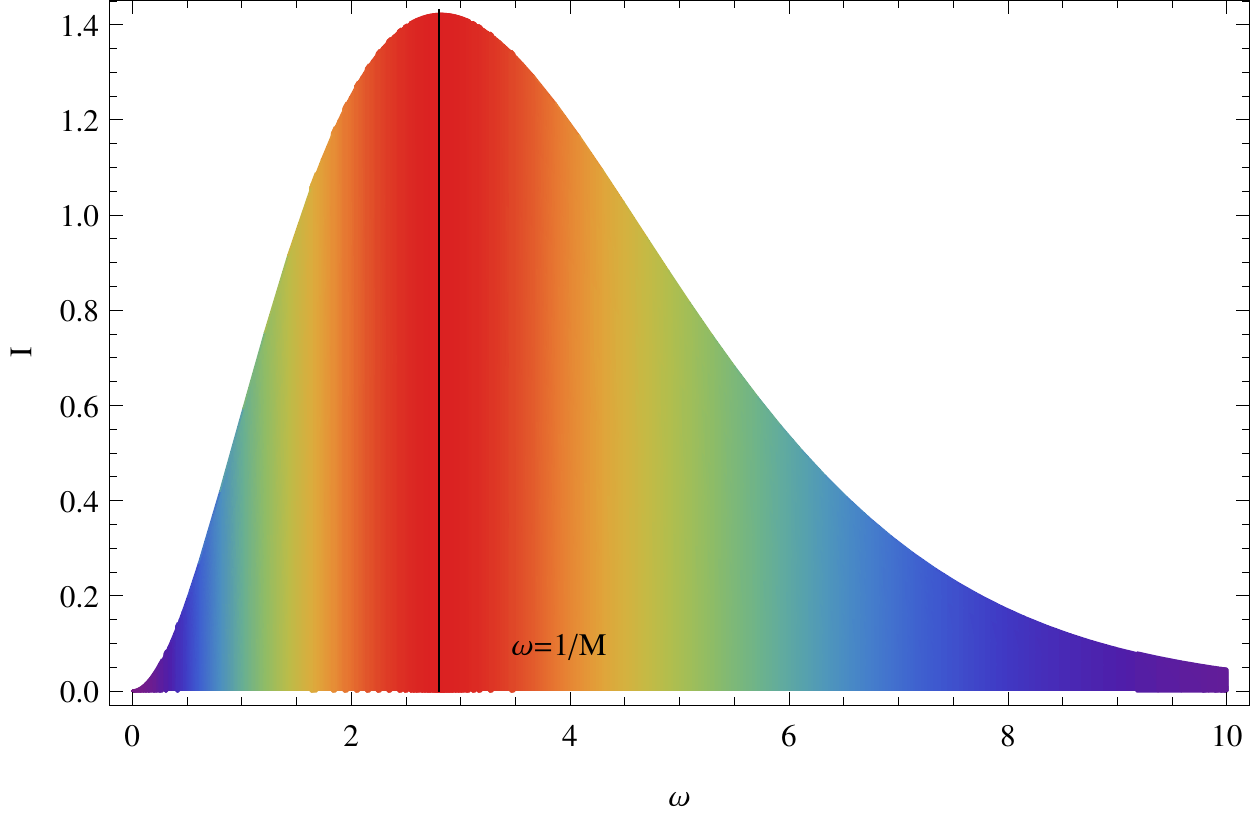}\hspace{2.0cm}
\includegraphics[scale=0.5]{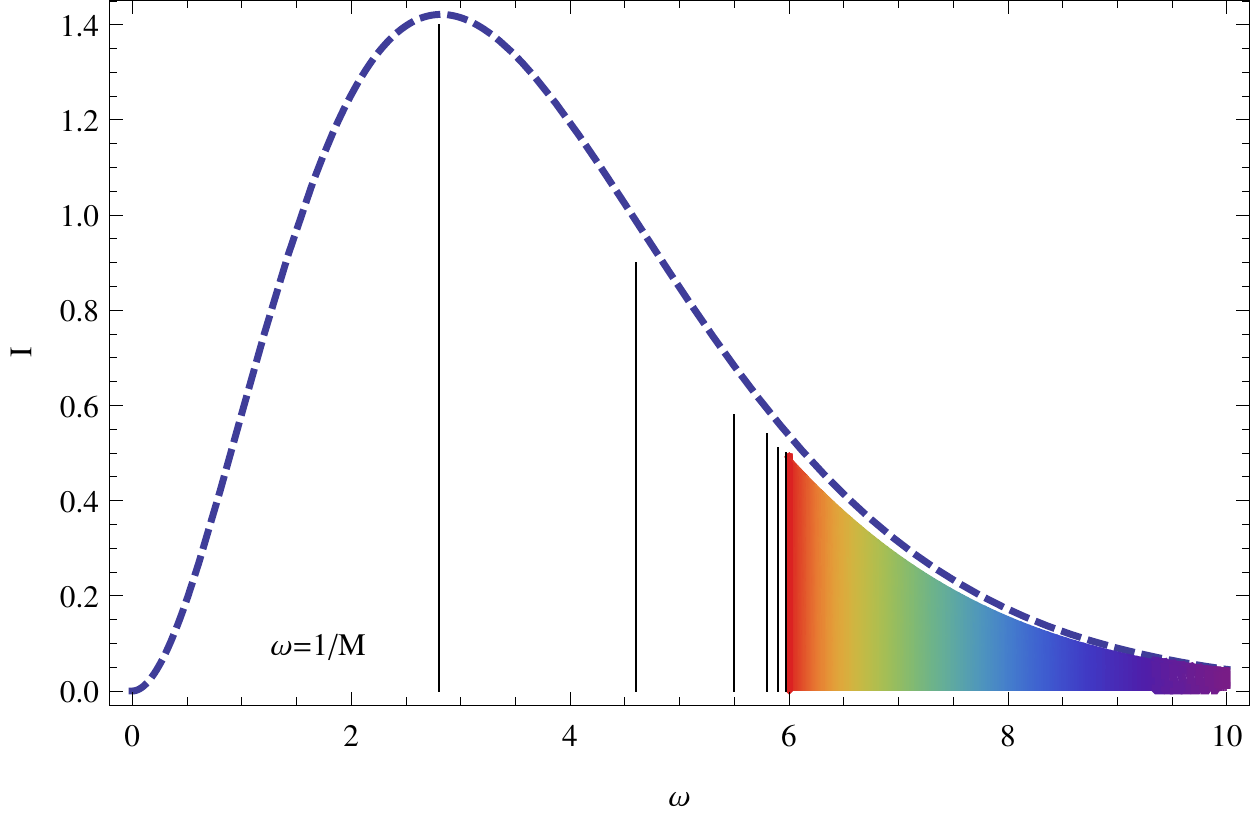}\\
\caption{The diagram in the left illustrates the semi-classical expectation of a thermal Hawking spectrum with the maximum amplitude at $\omega =1/M$ denoted by black line. While the one on the right depicts the scenario for the geometry quantized theories where there is no line present between $(0,1/M)$.}
\label{Penrose}
\end{figure}

A horizon is characterized by its macroscopic area $A$. The classical expectation of the horizon being a holographic screen corresponds to,
\begin{align}
S=A/4 = k_B \log \Omega \label{B-H relation}
\end{align}
where $\Omega$ is the number of micro-states corresponding to the macroscopic black hole configuration. This holographic relation will turn out to be a huge constraint on the quantized black hole geometry. Clearly, $\Omega$ has to be an integer, counting the number of micro-states. Let, microscopically, the area (of the horizon) gets quantized in a quantum theory of gravity as $A=\alpha (\sum _{j}n_{j}f(j))^{\beta}$, for an arbitrary $\beta$, with $\alpha$ being a parameter relating microscopic physics to macroscopic values. If the condition: $\sum_jn_j \gg1$ is satisfied, $A$ will be the macroscopic area of the hole. Macroscopic black hole configurations can get realized only with constraints: $\alpha =4 \log{K}; K\in \mathbb{Z}_+ $ and $(\sum _{j}n_{j}f(j))^{\beta}= I\in \mathbb{Z}_+$, where $\mathbb{Z}_{+}$ is the set of natural numbers. Owing to these two constraints, the Hawking spectra becomes non-thermal. Note that for $\beta =1$, the arguments of Bekenstein and Mukhanov turn alive again, this time 
as the {\it macroscopic} area gets integerly spaced. The arbitrariness in $f(j)$ or $\beta$ shows that scheme of geometry quantization can be as arbitrary as one wishes. Still, \textbf{in any case} the area-entropy relation dictates that the minimum mass change (i.e. the  minimum unit of energy extraction) to be \cite{Lochan:2015bha}
\begin{align}
\hbar \omega _{\rm min} \equiv \left(\delta M\right)_{\textrm{min}}=\delta \left(A^{1/2}\right)_{\textrm{min}}  \sim (I+1)^{1/2} - (I)^{1/2} \sim\frac{1}{M}. \label{Eqn1}
\end{align}
Hence the minimum frequency of quanta emitted by a macroscopic black hole corresponds to the characteristic frequency of the semi-classical thermal spectrum associated with the maximum amplitude. Thus there is \emph{no} transition line present at the fully quantum level, in the Hawking spectra for frequencies below $1/M$ (see \ref{Penrose}). To stress again, this result holds for arbitrary spectral behavior, i.e., any quantum gravity theory with quantized geometry can \emph{never} reproduce thermal Hawking spectrum. There is also an upper cut-off on frequencies of emission, which is the black hole mass $M$. Therefore, the most optimistic scenario would be when the emission is evenly spread between $1/M$ and $M$. Since there is no emission line between $(0,1/M)$, the emission spectrum has to channelize roughly $45\%$ of blackbody energy spectrum to high frequency modes. {\it Hence, a geometry quantized theory enhances the population of high frequency modes}, increasing the observational prospects. Existence of the minimum frequency as depicted in \ref{Eqn1} explicitly shows that a quantized geometry breaks the maximal entanglement of quanta of emitted radiation with those in the interior of black hole at macroscopic scales, making them {\it much more} information rich.
What could have we done to make the emission profile appear thermal, as in semi-classical physics? The answer, within the geometry quantized paradigm, is --- {\it absolutely nothing!} One may think of using higher curvature, higher dimensional gravity theories, improvised classical relations or using even more complicated quantum spectrum (even of different observables related to microscopic area in complicated manners), which went into the derivation of the discrete emission profile. However, simple reasoning, on the above mentioned lines, shows the robustness of our result in all such modifications \cite{Lochan:2015bha}. If the geometry is not quantized it is straightforward to show that there are $\exp(M^{2}/4)$ transition lines present in the regime $(0,1/M)$. Thus the infrared regime gets highly populated if black hole geometry is not quantized. This sparseness is indeed quantum geometric in origin. Another consideration, which we discuss below, drives this point home.
\paragraph{\it \textbf{Let's cool things down : A quantum route towards extremality --- Kerr black holes :}} 

For astrophysical ratifications, we need to consider Kerr black holes, characterized by mass $M$ and specific angular momentum $a=J/M$, in view of the discussions presented above\footnote{whatever holds for Kerr black hole is applicable to Reissner-Nordstr\"{o}m and Kerr-Newman black holes as well.}. This will cross check two conceptual points for us --- (a) Whether the minimum frequency bound still holds for geometry quantized theory and (b) is there any way of obtaining a dense spectrum astrophysically ? Kerr black hole being a solution of Einstein gravity, has the entropy related to area by the same relation $S=A/4$. However, the mass-area relation gets modified, since the angular momentum is non-zero, taking the form $A=4\pi r_{h}^{2}$, with $r_{h}=M+\sqrt{M^{2}-a^{2}}$.

As done previously, we will assume that black hole is a macroscopic object with some geometrical variable being quantized, such that its macroscopic area is given by $\alpha(\sum _{j}n_{j}f(j))^{\beta}$, with identical interpretation. From entropy-area relation one arrives at similar constraint, e.g., $(\sum _{j}n_{j}f(j))^{\beta}$ should be an integer. The transition of a Kerr black hole characterized by $(M,a)$ with microscopic identifier $\lbrace n_{j}\rbrace$ to another Kerr black hole characterized by $(M',a)$  \footnote{In any emission from rotating macroscopic black holes, it follows that change in the parameter $a$ is given by $\delta a\simeq -\eta \delta M$, with $\eta =a/M$. For $\eta <1$, i.e. non-extremal black hole, the black hole has a tendency to lose more mass, than the specific angular momentum $a$ through the quantum emission, i.e., $\delta a/\delta M \sim 0$, for macroscopic holes.} with microscopic counting $\lbrace n_{j}'\rbrace$ would be minimum if the integer shifts by unity. One can 
associate a maximum wavelength of emitted quanta $\lambda _{\rm max}$ as the inverse of the minimum frequency $\omega _{\rm min}$. For Kerr black hole one can define two more wavelengths --- (i) the Compton wavelength $\lambda _{c}$, which is numerically equal to the horizon radius and (ii) de-Broglie wavelength $\lambda _{T}$, obtained as the inverse of the semi-classical temperature of the black hole. It turns out that these quantities are related through a simple identity \cite{Chakraborty:2016},
\begin{align}\label{Eq_02}
\frac{\left(\lambda _{\rm max}\right)}{\left(\lambda _{\rm max,sch}\right)}=\frac{1}{16\pi}\left(\lambda _{c}\right)\left(\lambda _{T}\right)
\end{align}
where every $\lambda$ is in the unit of mass $M$ and $\lambda _{\rm max,sch}$ corresponds to the maximum wavelength for the non-rotating case. For Schwarzschild case the left hand side is unity, signaling the fact that both the Compton and de-Broglie wavelengths are similar, i.e., $\lambda _{c}\sim \lambda _{T}\sim M$, but for Kerr black hole, they become different. In the extremal limit, i.e., $M\rightarrow a$, the Compton wavelength tends to $M$, while the de-Broglie wavelength diverges. Thus, the mass of the black hole keeps decreasing, till the black hole becomes extremal and the minimum frequency vanishes, leading to \emph{dense} Hawking spectrum. Further as $\eta \sim 1$, the change in $a$ and the change in $M$ would turn out to be equal and the black hole stays put on the extremal value in subsequent emissions. Thus once the black hole has reached extremality, it will remain extremal under emission. This seemingly provides a quantum route towards extremality, which thermodynamically looked 
unachievable and a dense spectrum. This also refutes the ``box size" cut-off explanation of the discreteness of emission spectrum \cite{DiNunno:2008id,Bekenstein:1974jk,Kotwal:2002ch}. Importantly, dense or not, the spectrum  essentially remains non-thermal contrary to semi-classical expectation. Further, for Schwarzschild case, we know that by losing mass the black holes increase their temperature and evaporate more quickly, at least semi-classically, whereas for the Kerr case, we see that the natural drag via emission is towards extremality, i.e., towards the zero semi-classical temperature, {\it which is very non-trivial effect of a minimum mass gap resulting from geometry quantization.} 
\paragraph*{Acknowledgements} S.C. thanks CSIR, Government of India for providing SPM fellowship. The authors thank T. Padmanabhan for useful comments on an earlier draft of the manuscript.
\bibliography{Gravity_1_full,Gravity_2_partial}
\bibliographystyle{./utphys1}
\end{document}